\documentclass[12pt]{article}

\usepackage[pdftex]{graphicx}

\usepackage{amssymb,amsfonts,amsmath}

\author{Erzs\'ebet Ravasz$^{1,2}$, S. Gnanakaran$^3$ and Zolt\'an Toroczkai$^{1,4}$}

\title{\bf Network Structure of Protein Folding Pathways}
\date{}

\begin{document}

\maketitle

\begin{center}
\small $^1$Center for Nonlinear Studies, Theoretical Division, Los Alamos National Laboratory, MS B258, Los Alamos, New Mexico, 87545\\
\small $^2$Department of Medicine, Beth Israel Deaconess Medical Center, Harvard Medical School, 330 Brookline Avenue, RW 763, Boston, MA 02215\\
\small $^3$Theoretical Biology and Biophysics Group, Theoretical Division, Los Alamos National Laboratory, MS K710, Los Alamos, New Mexico, 87545\\
\small $^4$Department of Physics, University of Notre Dame, Notre Dame, Indiana, 46556\\

\end{center}

\vspace{0.5cm}

\begin{abstract} The classical approach to protein folding inspired by statistical mechanics avoids the high dimensional structure of the conformation space by using effective coordinates. Here we introduce a network approach to capture the statistical properties of the structure of conformation spaces. Conformations are represented as nodes of the network, while links are transitions via elementary rotations around a chemical bond. Self-avoidance of a polypeptide chain introduces degree correlations in the conformation network, which in turn lead to energy landscape correlations. Folding can be interpreted as a biased random walk on the conformation network. We show that the folding pathways along energy gradients organize themselves into scale free networks, thus explaining previous observations made via molecular dynamics simulations.  We also show that these energy landscape correlations are essential for recovering the observed connectivity exponent, which belongs to a different universality class than that of random energy models. In addition, we predict that the exponent and therefore the structure of the folding network fundamentally changes at high temperatures, as verified by our simulations on the AK peptide.

 \end{abstract}

\section{Introduction}

Packing problems, atomic clusters \cite{1_Doye}, polymers, and the ultimate building blocks of life, proteins, are characterized by high-dimensional conformation spaces littered with non-accessible regions induced by self-avoidance. Here we use a network \cite{2_Newman, 3_Albert} framework to study the conformation space (the collection of all accessible spatial conformations) of chain-like structures such as polymers and proteins \cite{1_Doye, 4_Scala, 5_Rao}. Conformations are represented as nodes of the network, while links are transitions via elementary rotations around a chemical bond. Folding can be interpreted as a biased random walk on this conformation network. This framework allows us to identify the key statistical features needed to recover generic properties of folding dynamics. In particular, it has been observed via Molecular Dynamics (MD) simulations on a number of peptides \cite{5_Rao} that folding networks are scale-free with an exponent of $-2$. First we observe that folding networks are a special case of gradient graphs \cite{6_Toroczkai, 7_Toroczkai}, which are induced by local gradients of a scalar field (conformational energy) associated with the nodes of a substrate graph (conformation network). We find that the scale-free property is a generic feature of gradient networks and thus in particular of protein folding networks. Second, we identify the statistical properties of the substrate graph and scalar (energy) field responsible for the $-2$ exponent and show that it is a consequence of correlations in the energy landscape. We anticipate that the methodology presented here and the some of the conclusions (such as the scale-free character of energy landscape networks) can be carried over to other conformation spaces as well, including atomic clusters \cite{1_Doye} and other packing problems.

	The spatial conformation of proteins can be described by the sequence of the backbone dihedral angles $( \Phi, \Psi )$ between consecutive peptide bond planes \cite{8_Ramachandran}. Although these angles are continuous variables, they are known to take on a few preferred values (typically 3 for each angle) corresponding to local minima of the torsional potential energy \cite{8_Ramachandran, 9_Levinthal}. This allows for a natural representation of conformations as nodes of a network \cite{5_Rao} (conformation network), with edges representing rotations from one preferred dihedral angle to another around a single chemical bond (elementary rotations). For an $n$-monomer protein the conformation network has on the order of $10^n$ nodes (distinct states), which attains astronomically large values even for short peptides. The immense size of conformation spaces was first pointed out by Levinthal  \cite{9_Levinthal} (known as the paradox of protein folding): searching at random for a particular state (such as the native state) would take the peptide longer than the age of the universe \cite{10_Wetlaufer}. Historically, this size issue has been avoided by projecting the conformation space onto one or two variables such as reaction coordinates or order parameters (e.g., the root mean square deviation of the atoms from their positions in the native state). Unfortunately, this leads to loss of information about the structure of conformation spaces, which are naturally high dimensional. Additionally, the choice of reaction coordinates often requires the knowledge of the native state. The network representation preserves the structure and dimensionality of conformation spaces, and at the same time creates a framework for a statistical description of their structural properties. Since statistical properties frequently obey scaling laws, working with smaller but statistically similar networks avoids the size issue. The approach presented here is based on this premise: {\it Generic features of folding dynamics are determined by statistical properties of the conformation network. }

\section{Results and Discussion}

\subsection{Conformation networks}
In spite of the wide diversity among proteins (as distinguished by their amino-acid sequence), we argue that their conformation networks share important statistical features. Namely, their degree distributions are scaled --- sharply peaked around their average (characteristic to homogeneous networks \cite{11_Erdos}), and they have the small-world property  \cite{12_Watts}. These features are actually generic to chain-like systems, as illustrated by a simple ball-chain model (BC) of $n +1$ balls connected by thin rods (Figure 1A and 1C). If there are $m$ relative angular positions between two consecutive rods (bonds), every conformation of the ball-chain can be represented as a sequence of integers $(i_1, i_2, \ldots, i_n)$, $i,j \in \{0, \ldots, m-1\}$. Assuming these $m$ positions can only be accessed sequentially (blue links in Figure 1A) the chain conformations naturally form an $n$-dimensional hypercube with $m$ states along each axis (Figure 1B). Certainly, this is a homogeneous network \cite{11_Erdos} with a binomial degree distribution (see Figure 1F and Supporting Information). In spite of itÕs lattice structure, this network is also small-world: the network size $N$ (node number) is exponential in the number of monomers ($N = m^n$), while its diameter is given by the largest Manhattan distance, $(m-1)\cdot n$, or $(m-1)\, \log_m N$: hence the logarithmic scaling characteristic to small-world networks \cite{12_Watts}.
\begin{figure}[!h]
\begin{center}
\includegraphics[width=0.95\textwidth]{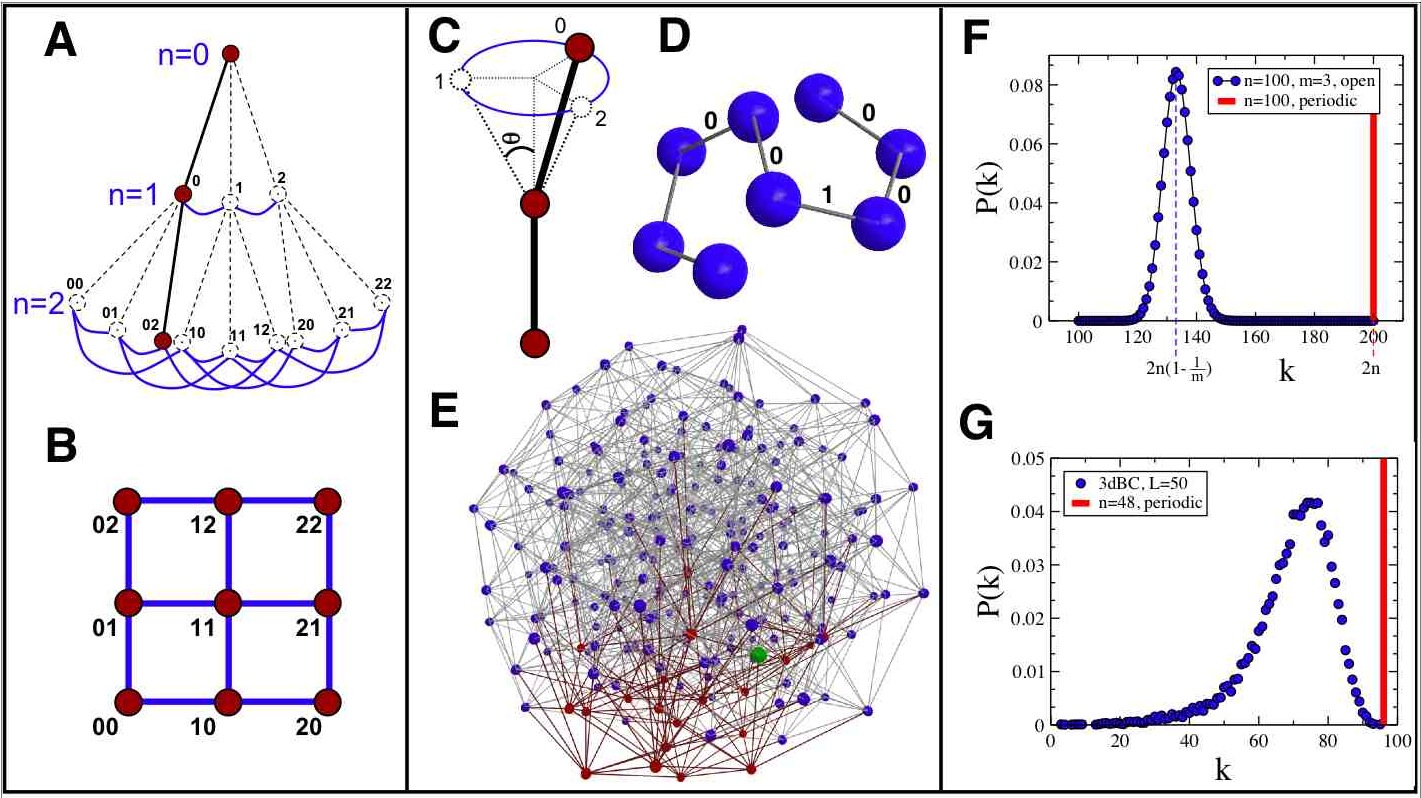}
\caption{{\bf Certain statistical properties of conformation networks can be inferred from simplified models.} Ball-Chain (BC) models in 2D and 3D (9).  {\bf(A)} 2d-BC model with $n=2$ monomers and $m=3$ allowed states between consecutive bonds. Blue lines represent allowed transitions Ð links of the conformation network.  {\bf(B)} 2d-BC conformation network redrawn as a lattice.   {\bf(C)} A simple 3d-BC model (see Supporting Information).   {\bf(D)} Conformation 00100, shown as a green node on E.  {\bf(E)} Conformation network of the 3d-BC model with $L = 7$ rods, $q = 75$ and $r = 0.25$. Sterically allowed conformations and transitions are shown with blue nodes and gray lines. Regions forbidden by hard-core exclusion are shown in red.   {\bf(F)} Degree distribution of the 100 dimensional conformation lattice with open (blue circles) and periodic (if state 0 and state m-1 are connected, red line) boundary conditions.   {\bf (G)} Degree distribution of the 3d-BC model (blue circles, $L = 50$, $q = 75$, $r = 0.25$, 20.000 sample points). The red line shows the degree distribution if sterical constraints are ignored.}\label{Fig1}
\end{center}
\end{figure}

Introducing the self-avoidance of protein chains into the BC model disrupts the perfect regularity of the conformation network: certain conformations (nodes) and transitions (links) are forbidden, i.e., pruned from the hypercube. This network resembles an $n$-dimensional Òswiss-cheeseÓ with holes representing the forbidden regions (Figure 1E). As a consequence, the degree distribution shifts towards lower degrees and broadens, however, it preserves its scaled character as shown in Figure 1G. The study by Scala {\it et al} of self-avoiding lattice polymers fixed at both ends \cite{4_Scala} recovers the same generic conformation network properties: binomial degree distribution and small-world nature.

\subsection{Folding Pathway Networks}
Recently, Rao and Caflisch \cite{5_Rao} have used Molecular Dynamics (MD) simulations to sample the conformation network of several 20-monomer peptides including beta3s (a designer peptide), its randomized heteropolymers, and homoglycine. Their result, however, presents a very different picture: the conformation networks of these peptides are all scale-free \cite{13_Barabasi}, with almost identical power-law tails for their degree distributions: $P(k) \sim k^{-2}$. We confirmed their results with MD simulations on the AK peptide \cite{14_Gnanakaran, 15_Paschek}. This suggests that the scale-free nature along with the $-2$  exponent is a universal property of protein conformation networks. Naturally two questions arise: 1) Why do simple ball-chain and lattice polymer models suggest a scaled network, while MD simulations of actual peptide chains indicate a scale-free structure? 2) What is behind the apparently universal character of the exponent $\gamma  = - 2$ ?  

To resolve the dilemma of question 1) we observe that conformation networks of chain models do not take into account the {\it potential energy} associated with different conformations (generated by the interactions between residues and between the chain and solvent). On the other hand, MD methods simulate conformational dynamics driven by energy differences between conformations. Since the conformation network enlists all sterically allowed states and transitions, the MD simulations will trace a path along the edges of this network. At $T = 0$ the path follows the local energy gradient, while at larger temperatures deviates from it according to Boltzmann statistics \cite{16_Scalley}. Hence, networks produced by MD simulations are temperature-dependent sub-graphs of the full conformation networks. 

\subsection{Gradient Networks}
One can characterize these MD graphs using the notion of the gradient network \cite{6_Toroczkai, 7_Toroczkai}: the collection of directed links that lead from every conformation (node) to their lowest energy neighbour. These directed links are organized into trees, dividing the conformation network into basins of local minima (Figure 2). At $T = 0$ the MD simulation paths follow the gradient links exclusively, while at higher temperatures they occasionally deviate from them, producing a "fattened" version of the gradient network. At very high temperatures the MD simulation performs an unbiased random walk on the conformation network.  
\begin{figure}[!h]
\begin{center}
\includegraphics[width=0.55\textwidth]{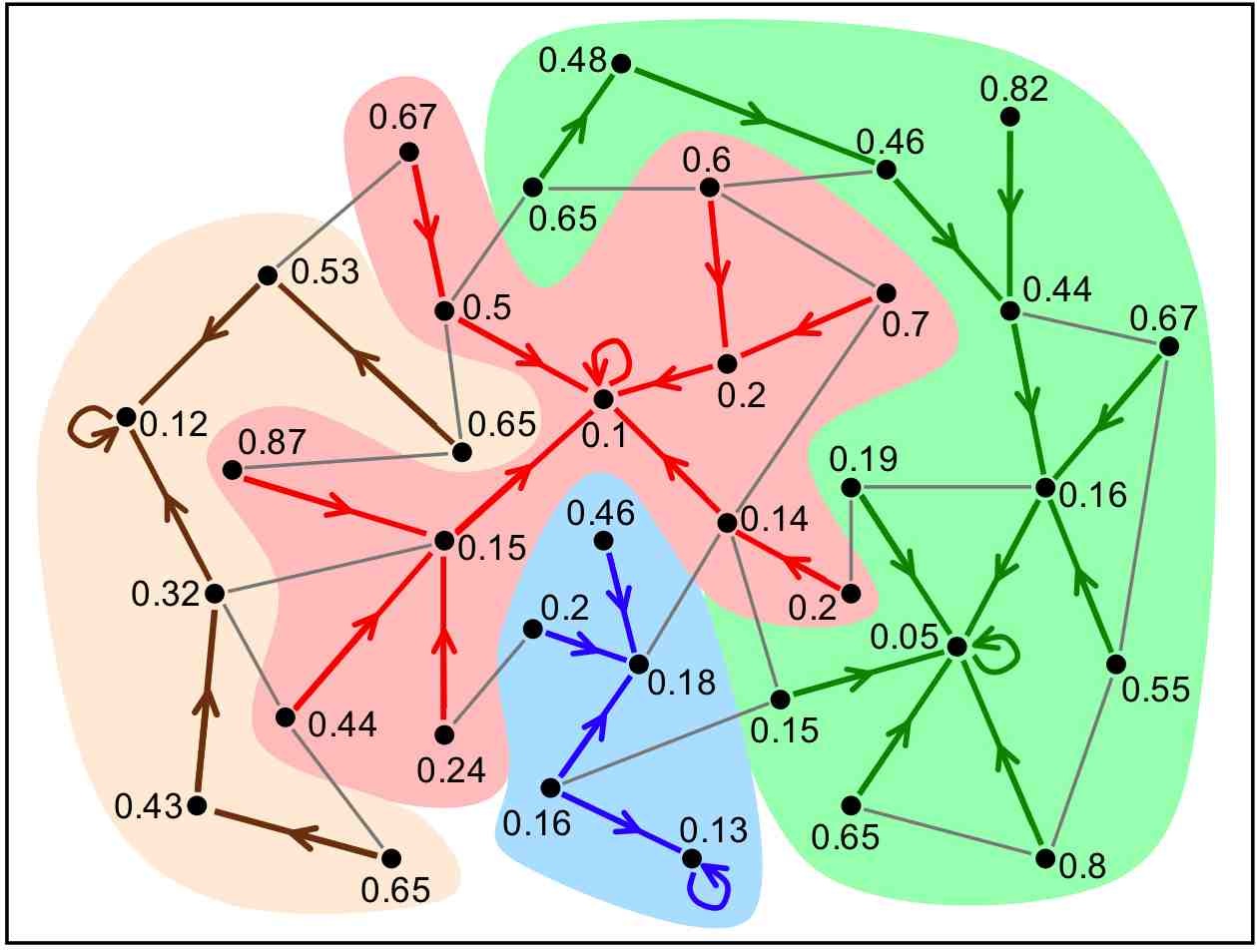}
\caption{{\bf The gradient network.} Gradient networks are directed subgraphs of a graph $G$ (the ÒsupportÓ network) generated by scalar fields associated with nodes on $G$. The links of the gradient network point from each node to the neighbour with the smallest scalar value (colored links). These links organize themselves into trees, which span the basins of local minima (the four basins shown with different colors; local minima are marked by self-loops).}
\label{Fig2}
\end{center}
\end{figure}

Gradient networks generated by random fields (a random scalar associated to each node) have been found to be scale-free \cite{6_Toroczkai, 7_Toroczkai} even if the supporting graph is scaled or homogeneous (e.g., Erd\H{o}s Ð R\'enyi graphs). The scale-free property can be analytically proven in the case of identical, independently distributed (i.i.d.) random scalars on scaled support graphs that do not contain loops of length 3 and 4 \cite{7_Toroczkai}. The gradient networks' scale-free nature, however, seems to be universal: we observed it for all types of substrate networks we investigated: regular tree, random tree, Erd\H{o}s-R\'enyi network, Small-World network, high dimensional regular lattice, $n$-torus lattice, the random geometric graph (Figs. S1, S2 in Supporting Information) and scale-free network (Barab\'asi-Albert, configuration model, etc. \cite{6_Toroczkai, 7_Toroczkai}). Since the MD simulations closely trace the gradient network, these observations explain the observed scale-free nature of the MD network, answering question 1). 

\subsection{Energy landscape correlations}
On scaled support graphs (such as conformation networks of heteropolymers) with i.i.d. random scalars (energies) distributed on them, the observed gradient degree exponent is always $\gamma  = -1$ (see Supporting Information, Figure S1 and S2) \cite{7_Toroczkai}, and not $-2$ as observed in MD simulations. Since the case of independently distributed random energies corresponds to the well-studied Random Energy Model of protein folding \cite{17_Bryngelson}, the discrepancy between the exponents shows the inadequacy of random energy models to characterize realistic folding landscapes \cite{18_Plotkin}. In order to deviate from the  Random Energy Model, one needs to uncover the correlations in the energy landscape responsible for the $-2$ exponent. 

First we observe that for proteins with effectively attractive interactions along the chain (a result of the interactions among the residues and the hydrophobic --- hydrophilic interactions with the surrounding solvent), the potential energy of tightly packed conformations (such as a native state) is lower on average, while for open and extended chain conformations it is larger (Figure 3A). For compact conformations, however, many elementary rotations are sterically forbidden and thus these represent nodes with low degree in the conformation network, while high degree nodes correspond to open chain conformations where virtually all of the elementary rotations are allowed. These observations show that {\it on average} the energy of a conformation ($\langle E \rangle$) is a monotonically increasing function of its degree ($k$) in the conformation network. For example, endowing the 3d-BC model with an attractive interaction between the balls given by a $-1/r^2$ potential leads to a monotonic behaviour of the $\langle E \rangle (k)$ function (Figure 3C). 
\begin{figure}[!h]
\begin{center}
\includegraphics[width=0.8\textwidth]{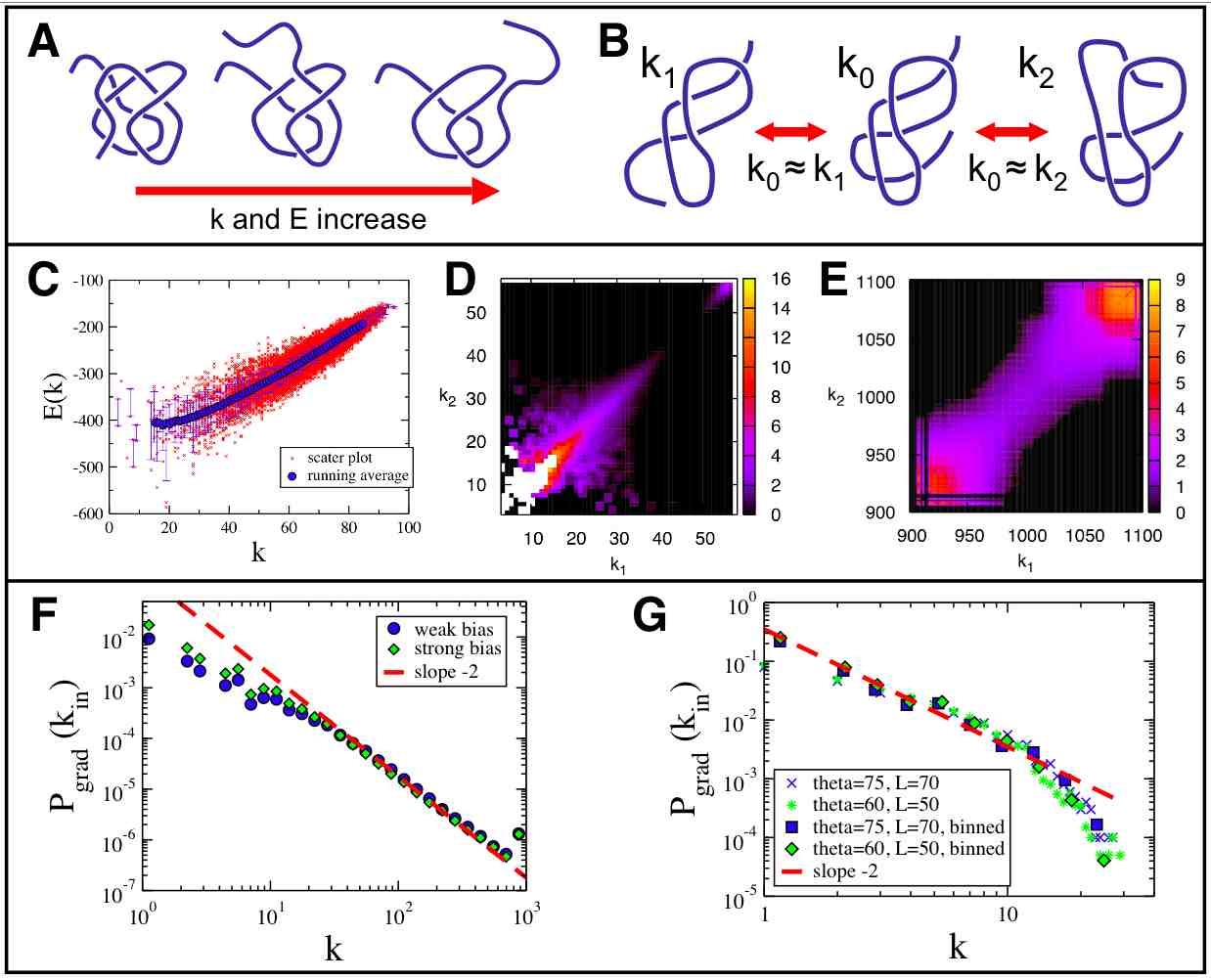}
\caption{{\bf Correlations in the structure of conformation networks.}  {\bf(A)} Correlations between degree and energy: tightly packed conformations have, on average, low degree as well as lower energy than loose ones.  {\bf(B)} Degree correlations or assortativity: tightly packed conformations with low degrees typically have neighbors with similar degrees.  {\bf(C)} Monotonically increasing $\langle E \rangle (k)$ function, as measured for the 3d-BC model with an attractive potential ($-1/r^2$) between the balls ($L = 50$, $q = 75$, $r = 0.25$, 20,000 sample points). {\bf(D)} Degree correlations in the 3d-BC model ($L = 30$, $q = 75$, $r = 0.25$, 20,000 sample points). Colors are proportional to the square of $K(k1,k2)$ (see Methods Section). {\bf(E)} Degree correlations in the RGG ($N = 30,000$, $\langle k \rangle = 1000$). {\bf(F)}  Gradient network in-degree distribution for RGG ($N = 30,000$,  $\langle k \rangle = 1000$, averaged over 200 realizations for strong, 500 for weak $\langle E \rangle (k)$  bias). Scalar values were drawn at random from a sliding interval, the center of which increased with node degree. {\bf (G)} Gradient network in-degree distribution for the 3d-BC model with $ -1/r^2$ potential between balls ($L = 70$, $q = 75$, $r = 0.25$ with 10,000 sample points and $L = 50$, $q = 60$, $r = 0.25$ with 20,000 sample points).}
\label{Fig3}
\end{center}
\end{figure}

However, the above energy-network correlation alone is not sufficient to produce the $-2$ scaling (as illustrated in the Supporting Information, Figure S3). One needs to include a second statistical ingredient, which is indeed characteristic to heteropolymer conformation networks as well, namely, degree assortativity \cite{19_Newman}. This means that connections between nodes with similar degree are highly probable, whereas connections between nodes with very different degrees are less likely. For conformation networks this holds naturally (see Figure 3D for the 3d-BC model), since one elementary rotation does not significantly unpack a compact (low degree) conformation or collapse an open (high degree) one (Figure 3B). 

	We expect that all scaled networks and associated scalar fields that share the two statistical properties above generate scale-free gradient networks with a $-2$ in-degree exponent. As an illustration we considered random geometric graphs (RGG) as substrate networks, obtained by connecting all pairs of randomly sprinkled $N$ points in the unit square that are within a prescribed distance $R$ (20). Similarly to the Erd\H{o}s-R\'enyi graphs \cite{11_Erdos}, these networks have a binomial degree distribution (thus scaled), however, unlike Erd\H{o}s-R\'enyi graphs they show degree assortativity \cite{19_Newman} (Figure 3E). Associating energy values that increase on average with node degree, one recovers the $\gamma = -2$ exponent for its gradient network (Figure 3F). For the 3d-BC model with $-1/r^2$ interactions the measured in-degree distribution of the generated gradient network is also consistent with the $\gamma = - 2$ exponent (Figure 3G, see Supporting Information for sampling issues). It is important to note that the $\gamma = -2$ exponent is a consequence of the {\em monotonic character} of the $\langle E \rangle(k)$ dependence, not on its specific form. The reason for this lies with the fact that gradient networks are only determined by the relative differences between the energies at the two ends of a link and not by their absolute values.

\subsection{Temperature dependence of the folding network}
Since the MD simulations trace a random walk on the conformation network biased by potential energy differences, we expect that this bias becomes gradually insignificant at larger temperatures and thus the deviations of the folding pathway network from the gradient network become more pronounced. As a consequence, the degree distribution of the MD pathway network should shift from a power-law scaling to a scaled form approaching the degree distribution of the full underlying conformation network (exponential tail).
We performed a series of MD simulations at increasing temperatures for the 20-monomer AK peptide \cite{14_Gnanakaran, 15_Paschek} (Figure 4A, also see Methods Section). As seen from Figure 4B, the degree distribution of the MD network shows a power-law decay with $\gamma = -2$ for lower temperatures, while at increasingly higher temperatures it transforms into a distribution with a fast decaying tail characteristic to homogeneous networks.
\begin{figure}[!h]
\begin{center}
\includegraphics[width=0.95\textwidth]{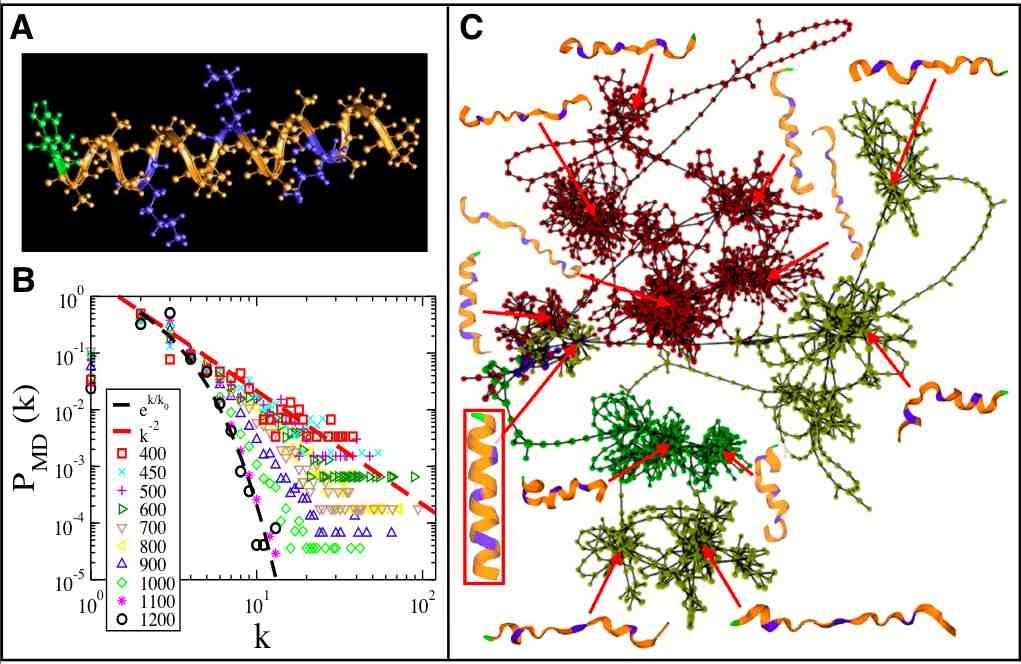}
\caption{{\bf MD simulations of the AK peptide.} {\bf(A)}  Helical conformation of the 20-monomer AK peptide (ALA: orange, LYS: blue, TYR: green). {\bf(B)}  Temperature dependence of the folding network. At low temperatures ($T = 400$) MD traces a network with a power-law degree distribution with exponent $-2$. However, increasing the temperature destroys this power-law: the network obtained at $T = 1200$ has a degree distribution with an exponential tail.  {\bf(C)}  Three different samples of the AK folding network at $T = 400$ (red, yellow and green), starting from the same conformation (helix shown in A).}
\label{Fig4}
\end{center}
\end{figure}

 Figure 4C shows the networks sampled by three short MD runs at $T = 400$ (starting from the same perfectly helical state but different random seeds) illustrating their degree heterogeneity and the basins of local minima that collect similar conformations. The helix partially melts in three different ways  (red, green, yellow networks) due to changes in initial conditions. The three  sampled  conformation networks have, nonetheless, robust statistical properties,
such as their degree distribution (red squares on Figure 4B).

\section{Conclusions}

 Here we have uncovered the microscopic origin of scale-free character and connectivity exponent for protein folding networks mapped from MD simulations of short peptides. Our approach provides a handle on the multi-dimensionality of the folding conformational energy landscape that is lost upon projecting onto low dimensions such as reaction coordinates. Further connection of this network with sequence information may prove valuable for identifying artificial sequences that are foldable, designing proteins with given requirements, and study conditions that may lead to miss-folding and aggregation. Furthermore, understanding the topology of the folding networks mapped by MD simulations will aid the development of faster computational algorithms to study the folding of large proteins.
 
\section{Methods}

\subsection{Measurement of degree correlations}

    Degree correlations in a network can be measured by comparing the number of links connecting a pair of nodes  $n(k_1,k_2)$ to it's value in an uncorrelated network, $n_0(k_1,k_2)$:
\begin{equation}
    K(k_1,k_2)=\frac{n(k_1,k_2)}{n_0(k_1,k_2)}-1= \frac{n(k_1,k_2)\,N}{(k_1+k_2)\,n(k_1)\,n(k_2)}-1,
\end{equation}
We used equation [1] to measure degree-degree correlations in the random geometric network (figure 3E).
For figure 3D, the 3d-BC model, we needed to measure degree correlations without constructing the entire network. We performed a random sampling of its topology by choosing a random node and mapping it's first and second neighborhood. In this case the available data is in the form of degree pairs
$k_\text{out}$ and $k_\text{in}$, where a distinction should be made between the randomly sampled nodes (with degree $k_\text{out}$) and their neighbors (with degrees $k_\text{in}$, which are not randomly sampled). Thus we used:
\begin{equation}
    K(k_\text{out},k_\text{out})=\frac{n(k_\text{out},k_\text{out})\,L}{k_\text{out}\,n(k_\text{out})\,n(k_\text{in})}-1,
\end{equation}
where $L$ is the number of sampled links (see Supporting Information).

\subsection{Molecular Dynamics Simulations of the AK peptide} 

$\,$ The model helical system considered in this study is a 20-residue
alanine-based peptide with 3 lysines (K), the AK peptide. The
sequence is A-A-A-A-K-A-A-A-A-K-A-A-A-A-K-A-A-A-A-Y. The AK peptide
was blocked with acetyl and amino groups at the N- and C- terminus,
respectively.
For this study, we have considered the AK peptide coupled to an
effective heat-bath instead of an explicit solvent. The modified
version of PARM94 force field of AMBER \cite{AMBER_1, AMBER_2} was used with an all atom representation for the AK
peptide. All bonds involving hydrogen atoms were constrained using
SHAKE \cite{SHAKE}. The initial conformation corresponded to the
fully helical AK peptide. Initial velocities were assigned randomly
to each atom from the Maxwell distribution for a given temperature.

Conformational preferences of the AK peptide are influenced by the
local environmental conditions and certainly by the surrounding
solvent. We have neglected the effect of solvent at this stage. 
We do not expect that the introduction of the solvent would influence the scale-free character of the folding pathways or their connectivity exponent $\gamma$, as long as the necessary correlations (as explained in the main text) are there. Indeed the MD simulations by Rao and Caflisch, which included the solvent, recover the same properties (scale-free and $\gamma = -2$) as our simulations on the AK peptide without a solvent.

 The output of the MD simulation is a list of conformation coordinates as a function of time. These coordinates allow us to measure the dihedral angles along the peptide backbone at every time-step and test for the observation by Ramachandran \cite{8_Ramachandran}, according to which the angular values in the $\Phi$--$\Psi$ plane are characterized by well-defined peaks. Since our simulations do not include solvents, it is important to check whether
the Ramachandran observation of preferred angular values still holds in this case.
For the AK peptide this would allow a good discretization of the $\Phi$--$\Psi$ plane. (For a standard discretization of amino-acid states of known proteins in their native state see the Protein Data Base \cite{PDB}. Their local secondary structure assignment is based on \cite{DSSP}). We performed 9 different temperature simulations of 0.2 nanoseconds each at temperatures ranging from 200K to 1,000K.  Conformations were sampled every 2 femtoseconds, the same as the MD time-step. Frequently visited values of $\Phi$--$\Psi$ organize in well-defined basins 
that correspond to the different local secondary structures the amino-acids are part of (see Supporting Information, figure S5). 
We divided the  $\Phi$--$\Psi$ plane in 7 domains, numbered them and discretized 
all  $\Phi$--$\Psi$ angles accordingly (our  $\Phi$--$\Psi$ domains, superimposed on contourplot representation of 
the angle distributions at different temperatures are shown in figure S6.

    The 9 simulations described above were also used to determine practical
    sampling rates at different temperatures. While at $T=1,000$ the system
    changes conformation in almost all $2\,$fs steps ($21,157$ different
    conformations were seen during one $100,000$ step run), at $T=200$ only
    15 different conformations were sampled in the same time. Thus we choose
    our sampling rate for every temperature such that the probability of two
    consecutive conformations being the same is $\simeq 0.45$ (the
    sampling rate only affects how often we record conformations, the
    time-step of the simulation itself is still $dt=2\,$fs, making low temperature runs longer).

     Figure 4B  was generated from simulations starting with
     an $\alpha$-helix state at 10 different temperatures ($T\in \{400,450,
     500,600,700,800,900,1000,1100,$ $1200\}$). The runs were ended when $100,000$
     sampled steps have been completed, with sampling steps of $Dt \in \{152,102,
     70,36,22,14,8,6,4,2\}\,$fs, respectively.   Figure 4C  was drawn using three $T=400$ runs of only  $10,000$
    sample steps each ($Dt=152\,$fs). They are all shorter than the time-scale on
    which the system would equilibrate and eventually reach a "native state"
    (assuming that it exists in the absence of a solvent).

\section*{Acknowledgments}
This work was supported in part by the Department of Energy under contract No. W-7405-ENG-36. 
\section*{Note}
For supporting information please e-mail: eregan@bidmc.harvard.edu.


\begin{thebibliography}{}

\bibitem{1_Doye} Doye J. P. K. (2002) {\it Phys Rev Lett} {\bf 88}, 238701.
\bibitem{2_Newman} Newman M. E. J. (2003)  {\it SIAM REV} {\bf  45}, 167.
\bibitem{3_Albert} Albert R, Barab\'asi A. L. (2002)  {\it Rev Mod Phys} {\bf  74}, 67.
\bibitem{4_Scala} Scala A, Amaral L. A. N, Barth\'el\'emy M. (2001)  {\it Europhys Lett} {\bf  55}, 594.
\bibitem{5_Rao} Rao F, Caflisch A. (2004)  {\it J Mol Biol} {\bf  342}, 299. 
\bibitem{6_Toroczkai} Toroczkai Z, Bassler K. E. (2004)  {\it Nature} {\bf  428}, 716.
\bibitem{7_Toroczkai} Toroczkai Z, Kozma B, Bassler K. E, Hengartner N. W, Korniss G. (2004) {\it  http://arxiv.org/abs/cond. mat/0408262}.
\bibitem{8_Ramachandran} Ramachandran G. N, Ramakrishnan C, Sasisekharan V. (1963) {\it  J Mol Biol} {\bf  7}, 95.
\bibitem{9_Levinthal} Levinthal C. (1969) in  {\it Mossbauer Spectroscopy in Biological Systems}, eds DeBrunner J. T. P, Munck E. (University of Illinois Press, Monticello).
\bibitem{10_Wetlaufer} Wetlaufer D. B. (1973)  {\it Proc Natl Acad Sci USA} {\bf  70}, 691
\bibitem{11_Erdos} Erd\H{o}s P, R\'enyi A. (1959)  {\it Publ Math. (Debrecen)} {\bf  6}, 290.
\bibitem{12_Watts} Watts D. J. (1999)  {\it Small Worlds: The Dynamics of Networks between Order and Randomness} (Princeton University Press, Princeton).
\bibitem{13_Barabasi} Barab\'asi A. L, Albert R. (1999)  {\it Science} {\bf  286}, 509.
\bibitem{14_Gnanakaran} Gnanakaran S, Hochstrasser R. M, Garcia A. E. (2004)  {\it Proc Natl Acad Sci USA} {\bf  101}, 9229.
\bibitem{15_Paschek} Paschek D, Gnanakaran S, Garcia A. E. (2005)  {\it Proc Natl Acad Sci USA} {\bf  102}, 6765.
\bibitem{16_Scalley} Scalley M, Baker D. (1997)  {\it Proc Natl Acad Sci USA} {\bf  94}, 10636.
\bibitem{17_Bryngelson} Bryngelson J. D, Wolynes P. G. (1987)  {\it Proc Natl Acad Sci USA} {\bf  21}, 7524.
\bibitem{18_Plotkin} Plotkin S. S,  Wang J, Wolynes P. G. (1996)  {\it Phys Rev E} {\bf  53}, 6271.
\bibitem{19_Newman} Newman M. E. J. (2002)  {\it Phys Rev Lett} {\bf  89}, 208701.
\bibitem{20_Dall} Dall J. C. (2002)  {\it Phys Rev E} {\bf  66}, 016121.

\bibitem{AMBER_1}  Cornell W. D., Cieplak P., Bayly C. I., Gould I. R., Merz Jr. K. M.,
				 Ferguson D. M., Spellmeyer D. C.,  Fox T., Caldwell J. W.,  Kollman P. A. (1995) {\it J Am Chem Soc} {\bf 117}, 5179.
\bibitem{AMBER_2} Garcia A.~E., Sanbonmatsu K., (2002) {\it Proc Natl Acad Sci USA} {\bf 99}, 2782.
\bibitem{SHAKE} Ryckaert J.~P., Ciccotti G., Berendsen H. J.~C., (1997) {\it J. Comput. Phys.} {\bf 23}, 327.
\bibitem{PDB} Berman H., Westbrook J., Feng Z., Gilliland G., Bhat T. N., 
			Weissig H.,  Shindyalov I. N.,  Bourne P. E., (2000) {\it Nucl Acids Res} {\bf 28}, 235.
\bibitem{DSSP} Kabsch W., Sander C., (1983) {\it Biopolymers} {\bf 22}, 2577.

\end{thebibliography}
\end{document}